%% file: main.tex
\documentclass[fleqn,usenatbib]{mnras}

\usepackage{newtxtext,newtxmath}

\usepackage[T1]{fontenc}

\DeclareRobustCommand{\VAN}[3]{#2}
\let\VANthebibliography\thebibliography
\def\thebibliography{\DeclareRobustCommand{\VAN}[3]{##3}\VANthebibliography}

\usepackage{graphicx}	
\usepackage{amsmath}	
\usepackage{longtable}
\usepackage{pdflscape}	
\usepackage{multirow}
\usepackage{lineno}
\usepackage{orcidlink}

\title[ASAS-SN neutrino follow-up]{ASAS-SN follow-up of IceCube high-energy neutrino alerts}

\author[J. Necker et al.]{
Jannis Necker \orcidlink{0000-0003-0280-7484}$^{1,2}$\thanks{E-mail: jannis.necker@desy.de},
Thomas de Jaeger \orcidlink{0000-0001-6069-1139}$^{3}$\thanks{E-mail: dejaeger@hawaii.edu},
Robert Stein \orcidlink{0000-0003-2434-0387}$^{1,2,4}$,
Anna Franckowiak \orcidlink{0000-0002-5605-2219} $^{1,5}$, 
\newauthor
Benjamin J. Shappee \orcidlink{0000-0003-4631-1149} $^{3}$,
Marek Kowalski \orcidlink{0000-0001-8594-8666} $^{1,2}$,
Christopher S. Kochanek$^{6, 7}$,
Krzysztof Z. Stanek$^{6, 7}$,
\newauthor
John~F.~Beacom \orcidlink{0000-0002-0005-2631} $^{8,6,7}$,
Dhvanil D. Desai \orcidlink{0000-0002-2164-859X} $^{3}$, 
Kyle Neumann \orcidlink{0000-0002-2701-8433} $^{6}$,
Tharindu Jayasinghe \orcidlink{0000-0002-6244-477X} $^{6}$,
\newauthor
T.~W.-S.~Holoien \orcidlink{0000-0001-9206-3460} $^{9}$\thanks{NHFP Einstein Fellow},
Todd A. Thompson$^{6, 7}$,
Simon Holmbo \orcidlink{0000-0002-3415-322X} $^{10}$
\\
$^{1}$Deutsches Elektronen Synchrotron DESY, Platanenallee 6, 15738 Zeuthen, Germany\\
$^{2}$Institut fur Physik, Humboldt-Universit\"at zu Berlin, D-12489 Berlin, Germany\\
$^{3}$Institute for Astronomy, University of Hawaii, 2680 Woodlawn Drive, Honolulu, HI 96822, USA.\\
$^{4}$Division of Physics, Mathematics, and Astronomy, California Institute of Technology, Pasadena, CA 91125, USA\\
$^{5}$Fakult\"at für Physik \& Astronomie, Ruhr-Universit\"at Bochum, D-44780 Bochum, Germany\\
$^{6}$ Department of Astronomy, The Ohio State University, 140 W. 18th Avenue, Columbus, OH 43210, USA.\\
$^{7}$ Center for Cosmology and AstroParticle Physics (CCAPP), The Ohio State University, 191 W. Woodruff Avenue, Columbus, OH 43210, USA.\\
$^{8}$ Department of Physics, The Ohio State University, 191 W. Woodruff Ave., Columbus, OH 43210, USA.\\
$^{9}$ The Observatories of the Carnegie Institution for Science, 813 Santa Barbara St., Pasadena, CA 91101, USA.\\
$^{10}$Department of Physics and Astronomy, Aarhus University, Ny Munkegade 120, DK-8000 Aarhus C, Denmark.\\
}

\date{Accepted XXX. Received YYY; in original form ZZZ}

\pubyear{2022}

\begin{document}
\label{firstpage}
\pagerange{\pageref{firstpage}--\pageref{lastpage}}
\maketitle

\begin{abstract}

We report on the search for optical counterparts to IceCube neutrino alerts released between April 2016 and August 2021 with the All-Sky Automated Survey for SuperNovae (ASAS-SN).
Despite the discovery of a diffuse astrophysical high-energy neutrino flux in 2013, the source of those neutrinos remains largely unknown. 
Since 2016, IceCube has published likely-astrophysical neutrinos as public realtime alerts. 
Through a combination of normal survey and triggered target-of-opportunity observations, ASAS-SN obtained images within 1 hour of the neutrino detection for 20\% (11) of all observable IceCube alerts and within one day for another 57\% (32). For all observable alerts, we obtained images within at least two weeks from the neutrino alert. ASAS-SN provides the only optical follow-up for about 17\% of IceCube's neutrino alerts.
We recover the two previously claimed counterparts to neutrino alerts, the flaring-blazar TXS 0506+056 and the tidal disruption event AT2019dsg.
We investigate the light curves of previously-detected transients in the alert footprints, but do not identify any further candidate neutrino sources. 
We also analysed the optical light curves of Fermi 4FGL sources coincident with high-energy neutrino alerts, but do not identify any contemporaneous flaring activity.
Finally, we derive constraints on the luminosity functions of neutrino sources for a range of assumed evolution models.

\end{abstract}




\section{Introduction}

Neutrinos are unique messengers from the high-energy Universe. Produced through interactions of high-energy cosmic rays with ambient matter and photon fields, they provide an unambiguous tracer of the sites of hadronic acceleration (see \citealp{ahlers2018} for a recent review). 
Following the discovery of a diffuse astrophysical neutrino flux by the IceCube collaboration \citep{Aartsen:2015rwa,2014PhRvL.113j1101A} 
there is now a major effort to identify their origin.
No significant clustering has yet been found within the neutrino data alone, but a search for neutrino clusters from known gamma-ray emitters found evidence for a correlation with the nearby Seyfert galaxy NGC\,1068 at the $2.9\sigma$ level \citep{2020PhRvL.124e1103A}. 

A complementary approach is to search directly for electromagnetic counterparts to individual high-energy neutrinos that have a high probability to be of astrophysical origin. Since 2016, the IceCube realtime program \citep{icecube17_realtime} has published their detections of such events through public realtime alerts and two candidate electromagnetic counterparts have since been identified at the $\sim 3\sigma$ level. In 2017, the gamma-ray blazar TXS 0506+056 was observed in spatial coincidence with a high-energy neutrino during a period of electromagnetic flaring \citep{IceCube:2018dnn}. A search for neutrino clustering from the same source revealed an additional neutrino flare in 2014-15 \citep{ice2018_txsflare}, during a period without any significant electromagnetic flaring activity \citep{Fermi-LAT:2019hte}. Theoretical models have confirmed that conditions in the source are consistent with the detection of one neutrino after accounting for Eddington bias \citep{Strotjohann:2018ufz}. However, explaining the ``orphan'' neutrino flare in 2014/15 proved to be difficult \citep{Reimer:2018vvw,Rodrigues:2018tku}.  
Statistically, the $\gamma$-ray blazar population contributes less than 27\% to the diffuse neutrino flux \citep{icecube17}.

In 2019, the Tidal Disruption Event (TDE) AT2019dsg was associated with a high-energy neutrino \citep{Stein:2020xhk}. Models have proposed various TDE neutrino production zones, including the wind, disk, or corona (see \citealp{Hayasaki2021jem} for a recent review) which are consistent with the detection of one high-energy neutrino. Radio observations of AT2019dsg confirm long-lived non-thermal emission from the source \citep{Stein:2020xhk, Cendes:2021bvp, mohan_21, matsumoto_21}, but generally challenge models relying on the presence of an on-axis relativistic jet \citep{winter21}. A population analysis constrained the contribution of TDEs to less than 39\% of the diffuse neutrino flux \citep{stein_19}. 
The coincidence of the TDE-like flare AT2019fdr with a high-energy neutrino \citep{reusch2021} strong dust echos in AT2019fdr and AT2019dsg motivated a search for similar events \citep{vanvelzen2021}. A correlation at $3.7\sigma$ level of such flares with high-energy neutrino alerts was found.
Taken together, these results suggest that the astrophysical neutrino flux has contributions from multiple source populations \citep{Bartos:2021tok}.
Other possible neutrino source populations 
include supernovae and gamma ray bursts.

Here we present the optical follow-up of 56 IceCube realtime alerts released between April 2016 and August 2021 with the All-Sky Automated Survey for SuperNovae (ASAS-SN; \citealp{Shappee14, Kochanek17b}).
ASAS-SN is a network of optical telescopes located around the globe that observes the visible sky daily. Its large field of view makes it well-suited for fast follow-up of IceCube alerts and enables 
searches for
transient neutrino counterparts.
In Section \ref{sec:alerts} we introduce the IceCube alert selection followed by the description of our optical follow-up. We present our analysis of possible counterparts in Section \ref{sec:followup}. 
We derive limits on neutrino source luminosity functions in Section \ref{sec:limits} and discuss our conclusions in Section \ref{sec:conclusions}.

\section{IceCube realtime alerts}
\label{sec:alerts}

\begin{table*}
    \begin{center}
        \begin{tabular}{c r r r r r r c}
            \hline
            \hline
            \multicolumn{1}{c}{\textbf{Event}} & \multicolumn{1}{c}{\textbf{R.A. (J2000)}} & \multicolumn{1}{c}{\textbf{Dec (J2000)}} & \multicolumn{1}{c}{\textbf{90\% area}}  &\multicolumn{1}{c}{\textbf{1d coverage}} & \multicolumn{1}{c}{\textbf{14d coverage}} & \multicolumn{1}{c}{\textbf{Signalness}}& \multicolumn{1}{c}{\textbf{Refs}}\\
            \multicolumn{1}{c}{}& \multicolumn{1}{c}{[deg]}&\multicolumn{1}{c}{[deg]}& \multicolumn{1}{c}{[sq. deg.]}& \multicolumn{1}{c}{[\%]} & \multicolumn{1}{c}{[\%]} & \multicolumn{1}{c}{[\%]} & \multicolumn{1}{c}{}\\
            \hline
            \input{latex/alert_table_observed}\\
            \hline
            \hline
            \multicolumn{8}{l}{$^{*}$For offline selected events no, signalness is given. Because they are promising events that were selected by hand, we assume a signalness of 50\%}
        \end{tabular}
        \caption{A summary of the 56 neutrino alerts followed up by ASAS-SN. In the first three column we list the name of the alert and its position. In columns three to six we give the 90\% rectangular localisation of the neutrino as sent out in the GCN and the fraction of this area covered by ASAS-SN in the first 24 hours and 14 days, respectively, after the neutrino arrival. Finally, we list the signalness of the event and the reference to the original IceCube GCN. For HESE events no signalness was given and we neglect these events to be conservative.}
        \label{tab:nu_alerts_observed}
    \end{center}
\end{table*}

\begin{table*}
    \centering
    \begin{tabular}{c r r r c c c c}
        \hline
        \hline
        \multicolumn{1}{c}{\textbf{Event}} & \multicolumn{1}{c}{\textbf{R.A. (J2000)}} & \multicolumn{1}{c}{\textbf{Dec (J2000)}} & \multicolumn{1}{c}{\textbf{90\% area}} &
        \multicolumn{1}{c}{\textbf{Reason}} &
        \multicolumn{1}{c}{\textbf{Refs}}\\
        \multicolumn{1}{c}{} & \multicolumn{1}{c}{[deg]}&\multicolumn{1}{c}{[deg]}&
        \multicolumn{1}{c}{[sq. deg.]} &\multicolumn{1}{c}{}& \multicolumn{1}{c}{}\\
        \hline
        \input{latex/alert_table_not_observed}\\
        \hline
        \hline
    \end{tabular}
    \caption{A summary of the 15 neutrino alerts that could not be observed by ASAS-SN. We list the event name and position in the first three columns. The area of the 90\% rectangular localisation of the neutrino is listed in column four and the reference to the IceCube GCN in column 5.}
    \label{tab:nu_alerts_not_observed}
\end{table*}

The IceCube neutrino observatory, located at the South Pole, is the world's largest neutrino telescope with an instrumented volume of one cubic kilometre \citep{IceCube:2016zyt}. The IceCube realtime program \citep{icecube17_realtime} has  released alerts since 2016 for individual high-energy (>100 TeV) neutrino events with a high probability to be of astrophysical origin. 
Initially, there were two alert streams: the \textit{Extremely-High Energy} (EHE) alerts and the \textit{High-Energy-Starting Events} (HESE) alerts. EHE events were reported with an estimate of the probability for the event to have an astrophysical origin, called \textit{signalness}. This quantity was not reported for the HESE alerts. The first alert was issued on 27th April 2016 \citep{ic160427a}.
To increase the alert rate and to reduce the retraction rate, these streams were replaced with a unified `Astrotrack' alert stream in 2019 \citep{icecube19_realtime}. All alerts are now assigned a signalness value, with the stream subdivided into Gold alerts (with a mean signalness of 50\%) and Bronze alerts (mean signalness of 30\%).

\begin{figure}
    \centering
    \includegraphics[width=0.45\textwidth]{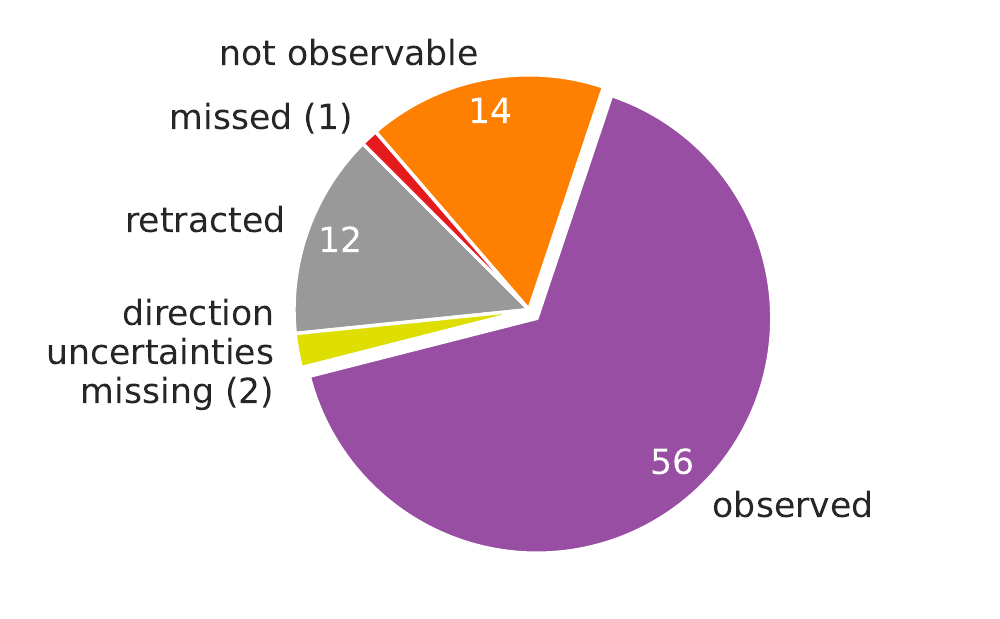}
    \caption{Statistics of ASAS-SN follow-up observations of the 85 IceCube alerts issued through to August 2021.}
    \label{fig:alerts_stats}
\end{figure}

A total of 85 alerts were issued before September 2021. Twelve were later retracted because they were consistent with atmospheric neutrino background events. For two alerts, IC190504A and IC200227A, IceCube was not able to estimate the uncertainty of the spatial localisation. Since the coverage of these alerts cannot be calculated, we exclude these two alerts from the subsequent analysis. The remaining 71 neutrino alerts were candidates for our follow-up program. 
A summary of the follow-up status of the alerts is shown in Figure \ref{fig:alerts_stats}. All IceCube neutrino alerts that could be followed up with ASAS-SN are listed in Table \ref{tab:nu_alerts_observed}. The ones that could not be observed are listed in Table \ref{tab:nu_alerts_not_observed}.

\section{Optical follow-up with ASAS-SN}
\label{sec:followup} 

\subsection{The All-Sky Automated Survey for Supernovae}

\begin{figure}
    \centering
    \includegraphics[width=0.45\textwidth]{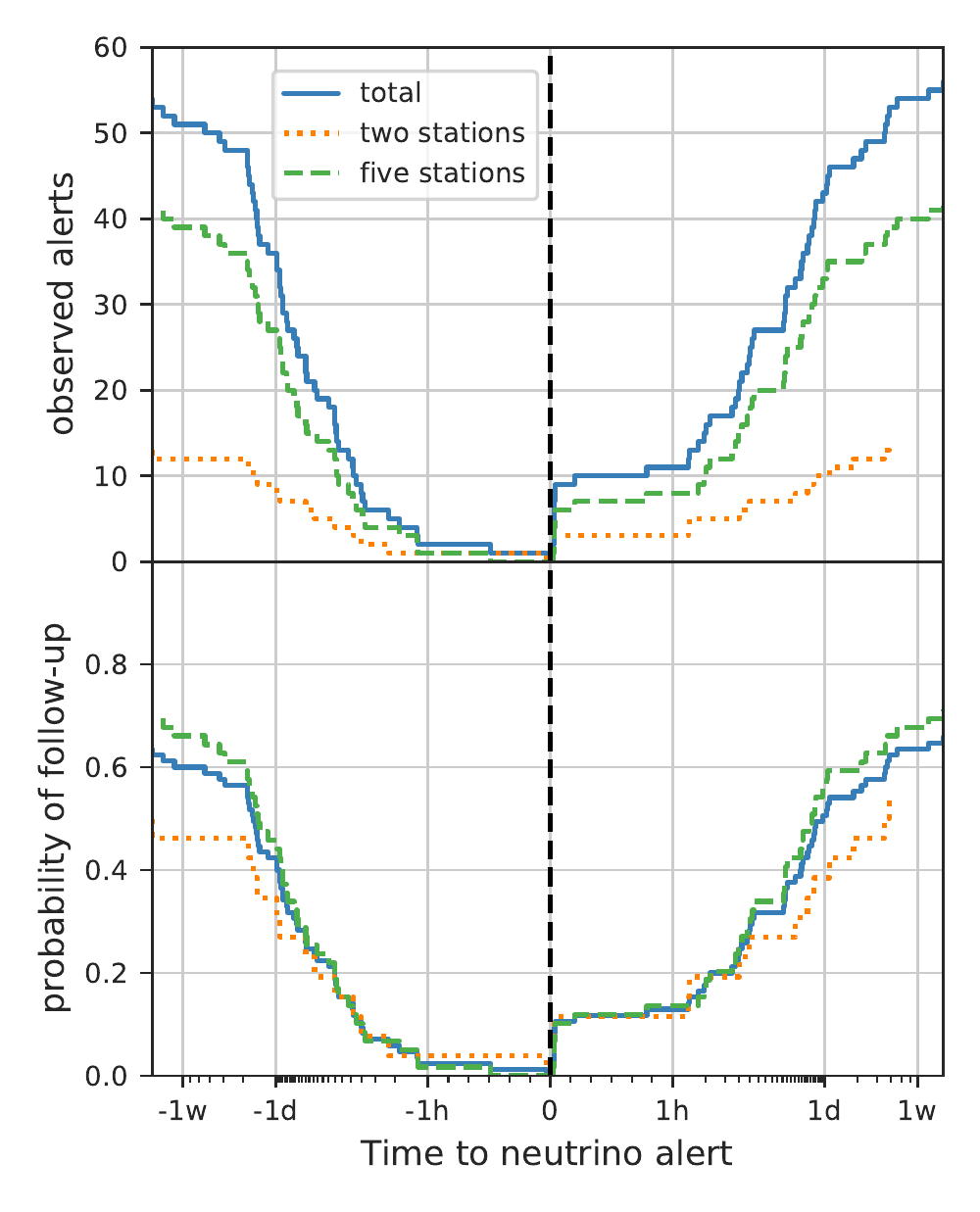}
    \caption{
    \textit{Top panel}: Number of events triggered by ASAS-SN within about two weeks of the IceCube alert. In less than one day, 43 events have been observed over a total of 56 triggered by ASAS-SN between 2016--2021.
    \textit{Bottom panel}: The mean probability of ASAS-SN observations for an IceCube alert.
    The boundary between the two station and the fully comissiond five station configuration is mid-2019.
    }
    \label{fig:alerts_24h}
\end{figure}

ASAS-SN is ideal to search for optical counterparts to external triggers such as IceCube neutrino alerts or gravitational-wave events, because it is the only ground-based survey to map the visible sky daily to a depth of $g = 18.5$ mag \citep{Shappee14, Kochanek17b}. ASAS-SN started late 2013 with its first unit, Brutus, located on Haleakala in Hawaii (USA). Shortly after, in 2014, ASAS-SN expanded with a second unit, Cassius, situated at Cerro Tololo International Observatory (CTIO) in Chile. Since late-2017, ASAS-SN is composed of five stations located in both hemispheres: the two original stations (Cassius and Brutus), Paczynski, also situated at CTIO in Chile, Leavitt at McDonald Observatory in Texas (USA), and finally Payne-Gaposchkin at the South African Astrophysical Observatory (SAAO) in Sutherland, South Africa. The two original units used a $V$-band filter until late-2018. The new units were installed using $g$-band filters and the two old units were switched from $V$ to $g$ after roughly a year of $V$- and $g$-band overlap. Each unit consists of four 14-cm aperture Nikon telephoto lenses, each with a 4.47 by 4.47-degree field of view. They are hosted by Las Cumbres Observatory (\citealt{Brown13}). 

The ASAS-SN survey has two modes of operation \citep{dejaeger21}: a normal survey operation mode and a Target-of-Opportunity mode (ToO) to get rapid imaging follow-up of multi-messenger alerts. During normal operations, each ASAS-SN field is observed with three dithered 90-second exposures with $\sim$15 seconds of overheads between each image, for a total of $\sim$315 seconds per field. For the ToO mode, we trigger immediately if there is a site that can observe the IceCube neutrino region. Thanks to the four sites, this is often the case. We obtain $\sim 15-20$ exposures for the pointing closest to the centre of the search region to go deeper and discover fainter candidates. All the images obtained from the ToO or the normal survey are processed and analysed in realtime by the standard ASAS-SN pipeline. A full description of the ASAS-SN optical counterpart search strategy can be found in \citet{dejaeger21}.

Prior to May 2017, only normal operation images were available. Once the ToO mode was implemented, we triggered on all the IceCube neutrino alerts and obtained images as soon as possible, in some cases within three minutes of the alert arrival time (IC190221A, IC190503A, IC200911A, IC201114A, IC201130A, IC210210A, and IC210811A). For one event (IC161103A), ASAS-SN was observing the respective localisation region as part of normal survey operations at the time of the neutrino arrival, resulting in images taken 105 seconds before and 2.5 sec after the alert arrival time. Since late 2017, there generally is a normal operations image ($\sim$18.5 mag) taken within a day if there are no weather or technical issues and the search region is not Sun or Moon constrained.
The bottom panel of Figure \ref{fig:alerts_24h} shows the cumulative distributions of observed events per year. 

To estimate the completeness of our observations, we draw lightcurves on random locations all over the sky. We inject simulated SN Ia lightcurves and test whether ASAS-SN would have detected the simulated supernova. For each lightcurve this is repeated 100 times. This gives a completeness down to 16.5 mag in V-Band and 17.5 mag in g-band, respectively. The analysis will be described in \citet[in prep.]{desai22}.

Fourteen neutrino alerts had a localisation too close to the Sun to be observed and one alert was missed due to the short observing window (less than 2 hours), leaving 56 that were followed up out of 71 real IceCube alerts. The top panel in Figure \ref{fig:alerts_24h} shows the cumulative number of events observed by ASAS-SN within about two weeks from the neutrino arrival, where the right side shows events observed after the neutrino arrival. Thanks to our strategy, we managed to observe 11 of the 56 triggered alerts in less than 1 hour (20\%) among which nine were observed in less than five minutes, another four in less than two hours (7\%), and 28 in less than one day (50\%). This illustrates our ability to promptly observe the majority of the IceCube alerts independent of the time or localisation. Finally, another thirteen events were observed  between 24 hours and two weeks (23\%; see Figure \ref{fig:alerts_24h}): four within two days, two in less than three days, four within four days, one in less than five days, and two within two weeks. Note that the longest delays in observation (IC200107A and IC201221A) were due to observability constraints or bad weather. So within at most two weeks, we observed all of the neutrino alerts that have not been retracted (12), have a well defined search region, and satisfy our observational restrictions: (1) the Sun is at least 12 degrees below the horizon, (2) the airmass is at most two, (3) the Hour Angle is at most five hours, and (4) the minimum distance to the Moon is larger than 20$^{\circ}$. 
The left side of the top panel in Figure \ref{fig:alerts_24h} shows the cumulative number of events that were serendipitously observed during routine observations. For 36 events we obtained images within 24 hours before the alert, which allows us to put better constraints on candidates. The localisation region of one alert (IC200530A) was observed about 30 minutes before the neutrino arrival and another one (IC161103A) was being observed at the time of neutrino arrival.
We also show the distributions for the periods before and after mid-2019. This marks the commissioning of the full five stations and the switch of the first stations to g-band (two stations and five stations in Figure \ref{fig:alerts_24h}, respectively). We calculate the probability of any event being observed by dividing the number of followed-up events by the number of neutrino alerts. The results are shown in the bottom panel of Figure \ref{fig:alerts_24h}. For any given neutrino alert ASAS-SN has a probability of about 60\% of obtaining observations. Most notably, the switch to the five station configuration significantly increased the probability of obtaining follow-up observations. For example it became 50\% more likeliy to obtain observations within one day.

Finally, it is worth noting that for 12 out of the 71 alerts (around 17\%) considered in this analysis, ASAS-SN observations are the only optical follow-up observation for the respective neutrino alert reported to the Gamma-ray Coordinates Network (GCN)\footnote{\url{https://gcn.gsfc.nasa.gov/selected.html}}.

\subsection{Possible Counterpart Classes to High-Energy Neutrinos}
\label{sec:counterpart_classes}

\begin{figure}
    \centering
    \includegraphics[width=0.5\textwidth]{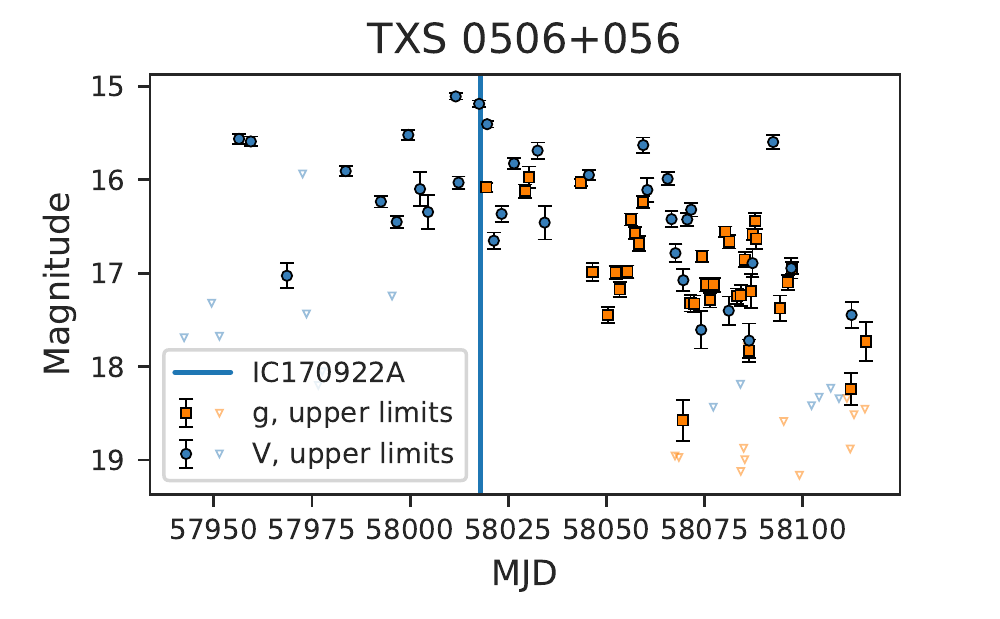}
    \includegraphics[width=0.5\textwidth]{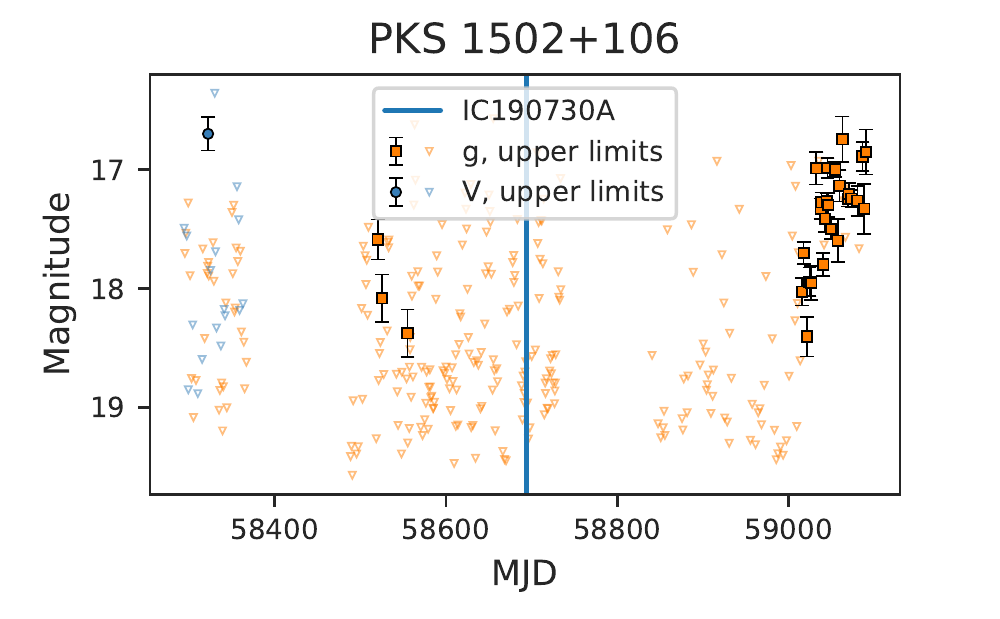}
    \caption{The ASAS-SN light curves of two blazars observed in spatial coincidence with high-energy neutrino alerts. We show $5 \sigma$ detections and upper limits. The date of the corresponding neutrino arrival is marked with a vertical line.}
    \label{fig:lightcurves_blazars}
\end{figure}

The challenge in identifying counterparts to high-energy neutrino events is that there are many possible neutrino source populations, each with different electromagnetic properties. Again, ASAS-SN's large field of view, fast response time, and archival data for the whole sky make it well suited for discovering transient counterparts to the IceCube neutrino events. 
The list of promising candidate source classes include:
\begin{itemize}
    \item \textbf{Supernovae with strong circumstellar material (CSM) interactions:} Models predict shock acceleration when the supernova ejecta interacts with the CSM \citep{murase11, murase2018, Zirakashvili16}. For sufficiently high-density CSM, strong interactions produce the narrow emission lines defining a Type IIn supernova \citep{schlegel1990, chugai1994}. 
    The shock can produce high-energy neutrinos for several years but for typical Type IIn conditions the flux is expected to have dropped by an order of magnitude after the first year.
    
    \item \textbf{Gamma-ray bursts (GRBs) and supernovae with relativistic jets:} Particle acceleration can occur inside the jet or at the shock where the jet interacts with the star's envelope \citep{meszaros01, senno16, ando2005}. This is true for both `successful' jets which escape the star and `choked' jets. In the first case the electromagnetic counterpart would be a stripped-envelope supernova with broad spectral features (a broad line Ic supernova, SN Ic-BL), and possibly a long GRB with an optical afterglow if the jet is aligned with the line of sight \citep{Woosley2006}. In the latter case the object would be a supernova Ic or Ib \citep{senno16}. In either case, a Type Ib/c SN with an explosion within a few days of the neutrino arrival is a compelling counterpart candidate, because the neutrino production is expected within tens of seconds of the core-collapse \citep{senno16}.
    
    \item \textbf{Tidal Disruption Events (TDEs):} TDEs have been proposed as high-energy neutrino sources, where neutrino production can occur in jets, outflows or winds, the accretion disk itself or the disk corona (see \citealp{Hayasaki2021jem} for a recent review). The TDE AT2019dsg 
    was observed in coincidence with a high-energy neutrino alert, where the neutrino arrived 150 days after the optical peak of the TDE \citep{Stein:2020xhk}. Another neutrino was observed about 300 days after the peak of the possible counterpart AT2019fdr \citep{reusch2021}. The timescale for non-thermal emission in TDEs can span several hundred days, so any active TDE coincident with a high-energy neutrino is interesting. This is especially true in the light of recently found indication of  correlation of high-energy neutrino alerts with TDE-like flares \citep{vanvelzen2021}.
    
    \item \textbf{Active Galactic Nuclei (AGN) Flares:} AGN flares may produce high-energy neutrinos by accelerating particles in accretion shocks \citep{stecker91}. This is especially true for blazar flares, where a jet points towards the Earth \citep{petropoulou15}. The blazar TXS0506+056 was observed in coincidence with a high-energy neutrino alert while it was in a flaring state \citep{IceCube:2018dnn}. Because these objects are numerous, we examined the ASAS-SN light curves of all Fermi 4FGL $\gamma$-ray detected blazars in the footprints of the neutrino alerts (see below and Figure \ref{fig:lightcurves_blazars}).
    
\end{itemize}

\subsection{Candidate Counterparts}
\label{sec:candidate_counterparts}

\begin{table*}
    \centering
    \begin{tabular}{c c c c c c}
    \hline
    \hline 
    \multicolumn{1}{c}{\textbf{Transient}} & \multicolumn{1}{c}{\textbf{ASAS-SN detection}} & \multicolumn{1}{c}{\textbf{IceCube alert}} & \multicolumn{1}{c}{\textbf{Alert epoch}} & \multicolumn{1}{c}{{$\mathbf{\Delta_{t}=t_{ASASSN}-t_{IceCube}}$}} & \multicolumn{1}{c}{\textbf{Transient type}} \\
    \multicolumn{1}{c}{} & \multicolumn{1}{c}{JD} & \multicolumn{1}{c}{} & \multicolumn{1}{c}{JD} & \multicolumn{1}{c}{days} & \multicolumn{1}{c}{}\\
    \hline
    ZTF18adicfwn (AT2020rng) &2459089.9 &IC210608A &2459373.7 &-284 &Unknown\\
    ATLAS19ljj (AT2019fxr) &2458634.9 &IC200410A &2458950.5 &-316 &Unknown\\
    ZTF19aapreis (AT2019dsg) &2458618.9 &IC191001A &2458758.3 &-139 &TDE\\
    ZTF19aadypig (SN~2019aah) &2458519.6 &IC191119A &2458806.5 &-287 &SN~II\\
    ASASSN-18mx (SN~2018coq) &2458286.1 &IC190619A &2458654.1 &-368 &SN~II\\
    ASASSN-17ot (AT2017hzv) &2458070.8 &IC180908A &2458370.3 &-300 &Unknown\\
    \hline
    \hline
    \end{tabular}
    \caption{An excerpt of Table \ref{tab:asassn_transients_long} of the transients that occur at most 500 days before the corresponding neutrino was detected, excluding spectroscopically-confirmed type Ia supernovae and CVs where neutrino emission is not expected.
    We give the name of the Transient and the Julian Date of its discovery in the first two columns. Columns three and four list the corresponding IceCube alert and the neutrino arrival time. In the last two column we give the difference between transient discovery and neutrino arrival and the transient type.}
    \label{tab:asassn_transients}
\end{table*}

\begin{figure*}
\begin{center}
    \includegraphics[width=0.95\textwidth]{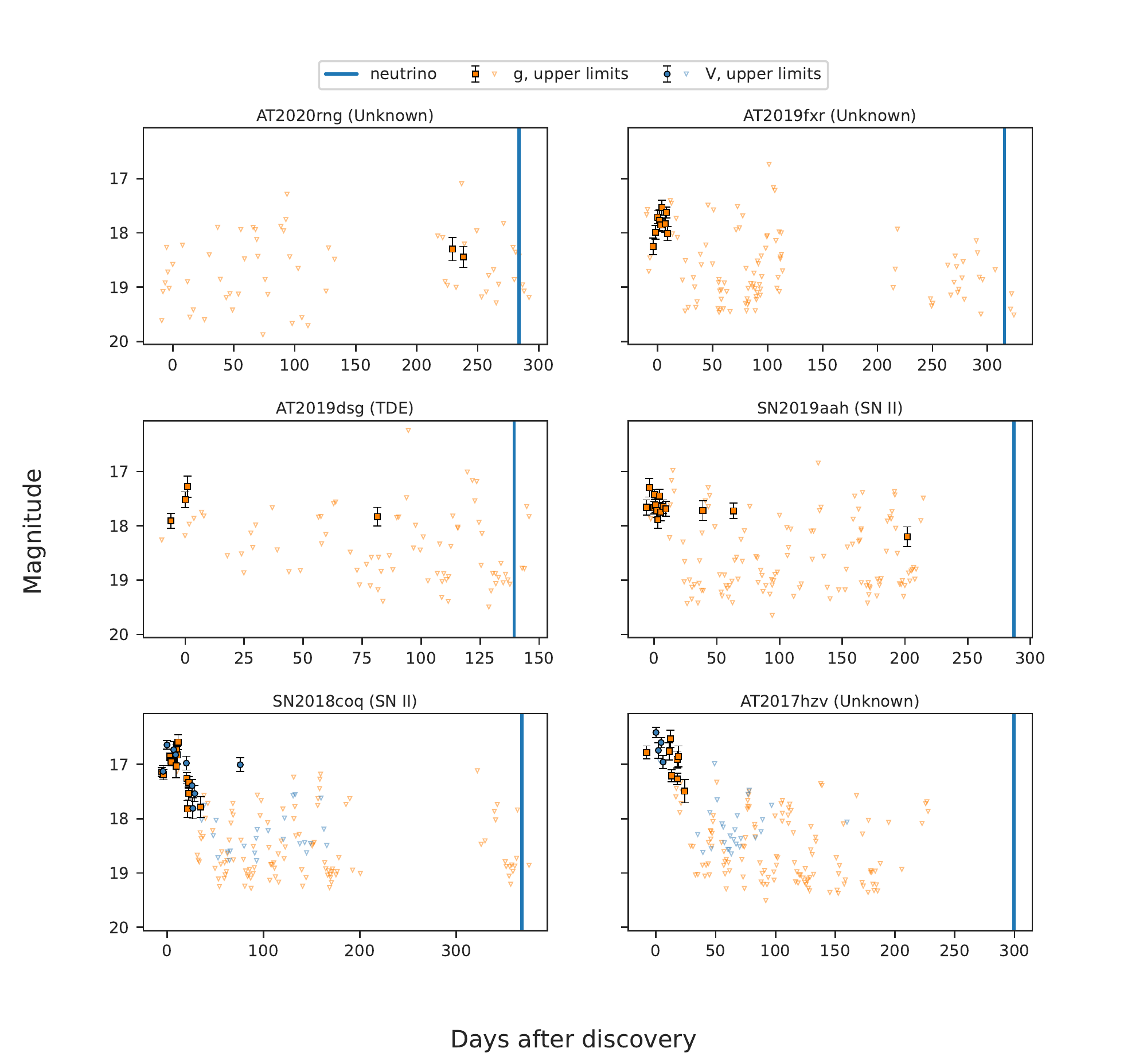}
    \caption{The ASAS-SN light curves for the transients found in the footprint of the IceCube neutrinos. We show $5 \sigma$ detections and upper limits as a function of the days after the discovery dates listed in Table \ref{tab:asassn_transients}. For AT2020rng, we used the Zwicky Transient forced-photometry service facility and show the $5\sigma$ detections in the r- and g-band (see Fig.~\ref{fig:at2020rng}). Vertical lines mark the time of the neutrino arrival.}
    \label{fig:candidate_lcs}
\end{center}
\end{figure*}

Table \ref{tab:asassn_transients} lists all transients identified by ASAS-SN in the 500 days prior to the neutrino arrival time excluding Type Ia SNe and dwarf novae (cataclysmic variables).
The list includes the paring of the TDE AT2019dsg and IC191001 \citep{Stein:2020xhk}. We do not detect the TDE AT2019fdr \citep{ic200530a_ztf} because it was too faint to trigger our transient detection pipeline.

The supernova SN~2019aah was spatially coincident with IC191119A. SN~2019aah was detected $\sim$300 days before the neutrino alert \citep{nordin_19aah} and was classified 30 days after the discovery as a Type II supernovae \citep{dahiwale2020_sn2019aah}. Its spectrum does not show narrow emission lines, so there is no evidence for a strong CSM interaction to produce neutrino emission. The emission is predicted to be strongest near the optical peak \citep{murase2019, Zirakashvili16}, so we conclude that SN 2019aah is unrelated to the neutrino.

SN~2018coq was spatially coincident with IC190619A. It is also a Type II SN \citep{cartier2018_sn2018coq}, discovered 370 days prior to the neutrino alert \citep{stanek18_sn18coq}.
Similar to SN~2019aah, its spectrum 13 days after the discovery does not show prominent narrow lines as a sign of CSM interaction. The supernova peaked even earlier relative to the neutrino than SN~2019aah, so SN~2018coq is unlikely related to IC190619A.

We find four neutrino-coincident events that could not be classified. All of them were first detected more than 280 days before the corresponding neutrino arrival.
AT2017hzv \citep{at2017hzv} and AT2019fxr \citep{at2019fxr} faded on the time scale of a few weeks and are not detectable at the time of the neutrino arrival.
The rapid fading suggest a supernova or AGN flare origin inconsistent with the neutrino arrival time which makes it unlikeliy they are associated with the corresponding neutrino.

For AT2020rng, we used the publicly available Zwicky Transient forced-photometry service \citep{masci2019}. We only find sporadic detections surrounded by upper limits (see Figure \ref{fig:at2020rng} in the Appendix). This, together with the relatively bright host galaxy with a mean g-band magnitude of 15.3 mag suggests that AT2020rng is a subtraction artefact rather than a physical transient.  

We also examined the ASAS-SN light curves of every Fermi 4FGL Blazar \citep{fermi2020_4fgl, fermi2020_dr2} within the footprint of a neutrino alert. We do not find any flaring activity coincident with the arrival of the corresponding neutrino, except for the previously-reported ASAS-SN observations of TXS 0506+056 \citep{IceCube:2018dnn}. This light curve is shown in the top panel of Figure \ref{fig:lightcurves_blazars}, with the source exhibiting an optical flare at the time of the neutrino detection. 

The neutrino IC190730A was observed in spatial coincidence with the Flat Spectrum Radio Quasar (FSRQ) PKS 1502+106 \citep{ic190730a,2020ApJ...893..162F} and the ASAS-SN light curve for this object is shown in the lower panel of Figure \ref{fig:lightcurves_blazars}. We confirm that the blazar was in a low optical state at the time of the neutrino arrival, as reported by \cite{ic190730a_ztf} and \cite{ic190730a_goto}. Time dependent radiation modeling found that the detection of a high-energy neutrino from this source is consistent with its multi-wavelength properties \citep{rodrigues_21}. 

\section{Limits}
\label{sec:limits}

\begin{figure}
    \centering
    \includegraphics[width=0.45\textwidth]{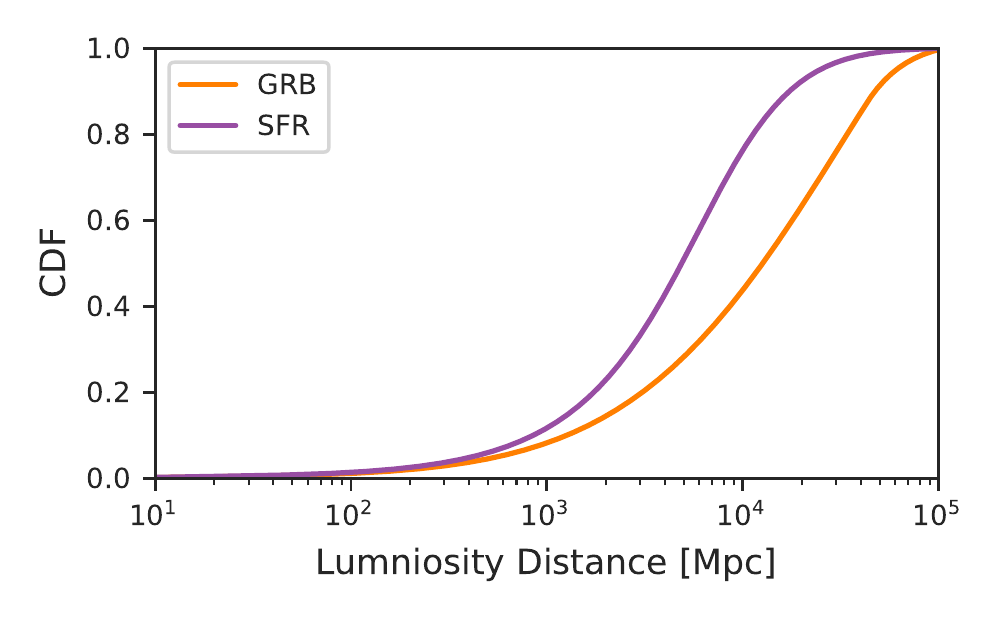}
    \caption{The relative cumulative neutrino flux at earth of neutrino source populations with a GRB-like and a SFR-like density evolution.}
    \label{fig:neutrino_cdfs}
\end{figure}

\begin{figure*}
    \centering
    \includegraphics[width=0.95\textwidth]{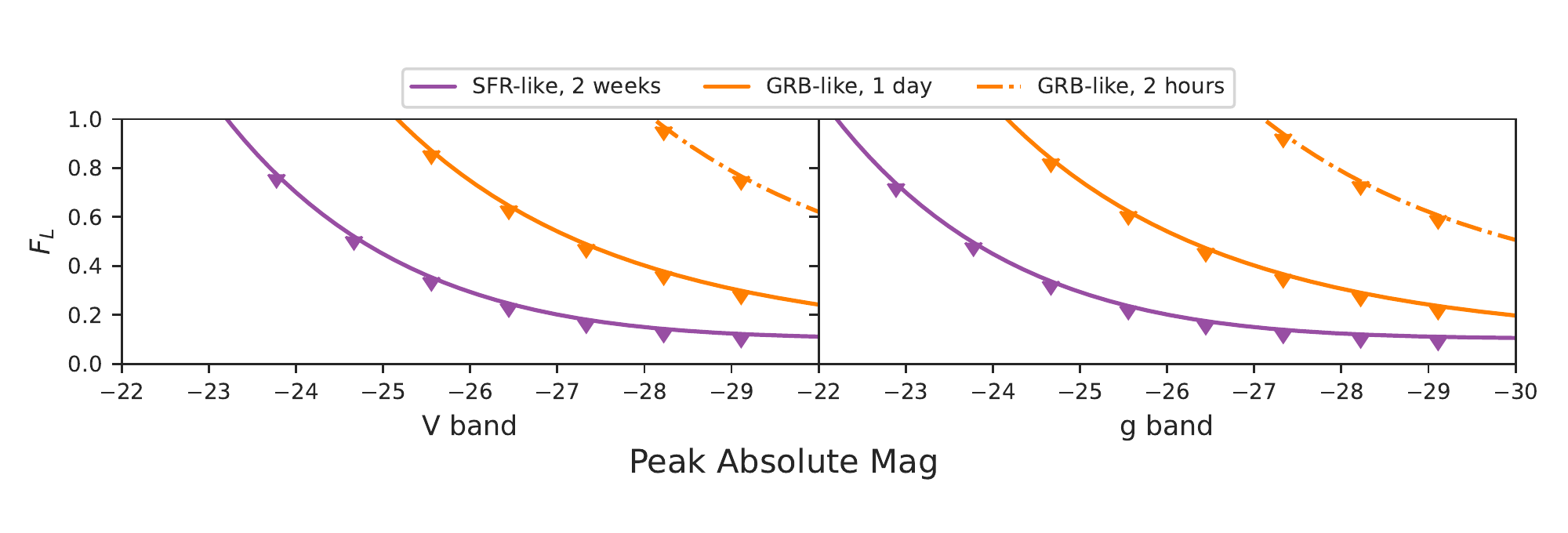}
    \caption{Constraints on the fraction $F_{\mathrm{L}}$ of a neutrino source population as a function of the intrinsic source Peak Absolute Magnitude.
    }
    \label{fig:limits}
\end{figure*}

While we do not find any new candidate counterpart transients in our follow-up campaign, we can use the non-detections to derive limits on neutrino source luminosity functions following the method of \citet[in prep.]{stein21b}. Because we recover the two pre-existing source candidates (TXS 0506+056 and AT2019dsg), these non-detection limits do not apply to blazars or TDEs. 

For an astrophysical neutrino with an electromagnetic counterpart, we can calculate the probability of detecting the counterpart based on the percentage of the neutrino localisation that was observed by ASAS-SN. For each neutrino this fraction is listed in Table \ref{tab:nu_alerts_observed} for one and fourteen days after the neutrino arrival. 
The probability of detecting a counterpart also depends on the probability for the neutrino to be of astrophysical origin. This is given by IceCube as the \textit{signalness} (see Section \ref{sec:alerts} and Table \ref{tab:nu_alerts_observed}).
We will assume in the following that we would have observed a transient if it had reached 18.5 mag and adapt it as the limiting magnitude of our program. 

At a 90\% confidence level we can constrain the fraction of neutrino sources above our limiting magnitude to be no more than 39.3\% and 15.3\% for fast transients which reach their peak within two hours and one day, respectively. For transients that peak within fourteen days, the fraction is 10.3\%. These constraints refer to the visibility of the transients and do not include any physical properties of the source classes.

To constrain physical populations of candidate neutrino sources we have to assume a rate $\dot{\rho}(z)$ at which the transients occur as a function of redshift $z$. We consider a GRB-like \citep{lien14_grbrate} and a star formation rate (SFR) like \citep{strolger15_sfrrate} source evolution. Because the optical afterglow of a GRB rapidly fades on the timescale of a few days \citep{Kann:2007cc}, we use the two hour coverage fraction of 15.3\%. Interacting supernovae typically rise on a timescale of at least two weeks \citep{nyholm2020} so we use the 39.3\% constraint for our coverage after 14 days.
The cumulative neutrino fluxes at earth from these populations as calculated with \texttt{flarestack} \citep{flarestack} are shown in Figure \ref{fig:neutrino_cdfs}. 

Assuming an absolute magnitude for the transient, we can compute the luminosity distance at which the transient would be at the apparent magnitude to which our follow-up program is complete. As a conservative choice we use the magnitude limit derived for the Type Ia SNe in ASAS-SN (see Section \ref{sec:followup}). Using the source evolutions from Figure \ref{fig:neutrino_cdfs}, we derive the neutrino flux that would arise in this volume from the corresponding neutrino source population. Given our limits on the fraction of the population above our limiting magnitude we can convert this into constraints on the fraction of sources $F_{\mathrm{L}}$ above the source absolute magnitude as shown in Figure \ref{fig:limits}.

These results are not yet constraining for typical supernovae with absolute magnitudes up to around $-21.5$. However, we can constrain the luminosity function of a neutrino source population with a GRB-like source evolution to produce counterparts that are below $-27$ magnitude in V-band about 54\% of the cases and in g-band about 40\% of the cases, one day after the neutrino arrival. This is the first such constraint on this timescale which is thanks to the high observation cadance and rapid follow-up of ASAS-SN.
\\

\section{Conclusions}
\label{sec:conclusions}
We presented the ASAS-SN optical follow-up program for IceCube high-energy, astrophysical neutrino candidates. We observed the 90\% localisation region of 56 alerts over the period from April 2016 until August 2021. Eleven of these alerts were covered within one hour after their detection. After 1 day we had observed 43 events and after two weeks we had observed the localisation regions for all 56 alerts to a limiting magnitude of $\sim 18.5$. For 12 events (around 17\%), this is the only optical follow-up. We did not detect any new coincident transients in our analysis, but we did recover the associations with the blazar TXS 05056+056 and the TDE AT2019dsg.

We find additional transients that we disfavour as 
counterparts of the corresponding neutrino. Given the non-detection of any transient counterpart in our search we derive upper limits on the luminosity function of different possible transient neutrino source populations.

Assuming the IceCube alert stream does not change, we can expect about 20 neutrino alerts per year. If our average coverage (18\% after two hours and 94\% after 14 days) does not change, we can set limits that are twice as strict on GRBs in 3.5 years and on CCSNe in 3 years, respectively.

The planned extension of IceCube, called IceCube-Gen2, is expected to increase the event rate significantly and improve the spatial resolution of through-going tracks \citep{2021icecube_gen2}. This will allow us to followup more neutrino alerts and cover a higher percentage of the smaller neutrino localisation area leading to an improved sensitivity to detect optical counterparts.

\section*{Acknowledgements}

J.N. was supported by the Helmholtz Weizmann Research School on Multimessenger Astronomy, funded through the Initiative and Networking Fund of the Helmholtz Association, DESY, the Weizmann Institute, the Humboldt University of Berlin, and the University of Potsdam.  Support for T.d.J. has been provided by NSF grants AST-1908952 and AST-1911074.  R.S. and A.F. were supported by the Initiative and Networking Fund of the Helmholtz Association, Deutsches Elektronen Synchrotron (DESY). B.J.S. is supported by NSF grants AST-1907570, AST-1908952, AST-1920392, and AST-1911074. CSK and KZS are supported by NSF grants AST-1814440 and AST-1908570. J.F.B. is supported by National Science Foundation grant No.\ PHY-2012955.  Support for TW-SH was provided by NASA through the NASA Hubble Fellowship grant HST-HF2-51458.001-A awarded by the Space Telescope Science Institute, which is operated by the Association of Universities for Research in Astronomy, Inc., for NASA, under contract NAS5-265. 

ASAS-SN is funded in part by the Gordon and Betty Moore Foundation through grants GBMF5490 and GBMF10501 to the Ohio State University, NSF grant AST-1908570, the Mt. Cuba Astronomical Foundation, the Center for Cosmology and AstroParticle Physics (CCAPP) at OSU, the Chinese Academy of Sciences South America Center for Astronomy (CAS-SACA), and the Villum Fonden (Denmark). Development of ASAS-SN has been supported by NSF grant AST-0908816, the Center for Cosmology and AstroParticle Physics at the Ohio State University, the Mt. Cuba Astronomical Foundation, and by George Skestos. Some of the results in this paper have been derived using the healpy and HEALPix packages.

The ZTF forced-photometry service was funded under the Heising-Simons Foundation grant \#12540303 (PI: Graham).

\section*{Data Availability}


All information about the IceCube neutrino alerts that we used are publicly available and can be accessed via the GCN archive for GOLD and BRONZE (\url{https://gcn.gsfc.nasa.gov/amon_icecube_gold_bronze_events.html}), the HESE (\url{https://gcn.gsfc.nasa.gov/amon_hese_events.html}) and EHE (\url{https://gcn.gsfc.nasa.gov/amon_ehe_events.html}) events.
The ZTF forced-photometry service is publicly available (see \url{http://web.ipac.caltech.edu/staff/fmasci/ztf/forcedphot.pdf} for a description of the access).



\bibliographystyle{mnras}
\bibliography{main} 




\appendix

\onecolumn

\section{}\label{AppendixA}
In Figure \ref{fig:at2020rng} we show $5 \sigma$ detections and upper limits for AT2020rng by ASAS-SN and the Zwicky Transient Facility forced-photometry service.
\begin{center}
\includegraphics[width=0.95\textwidth]{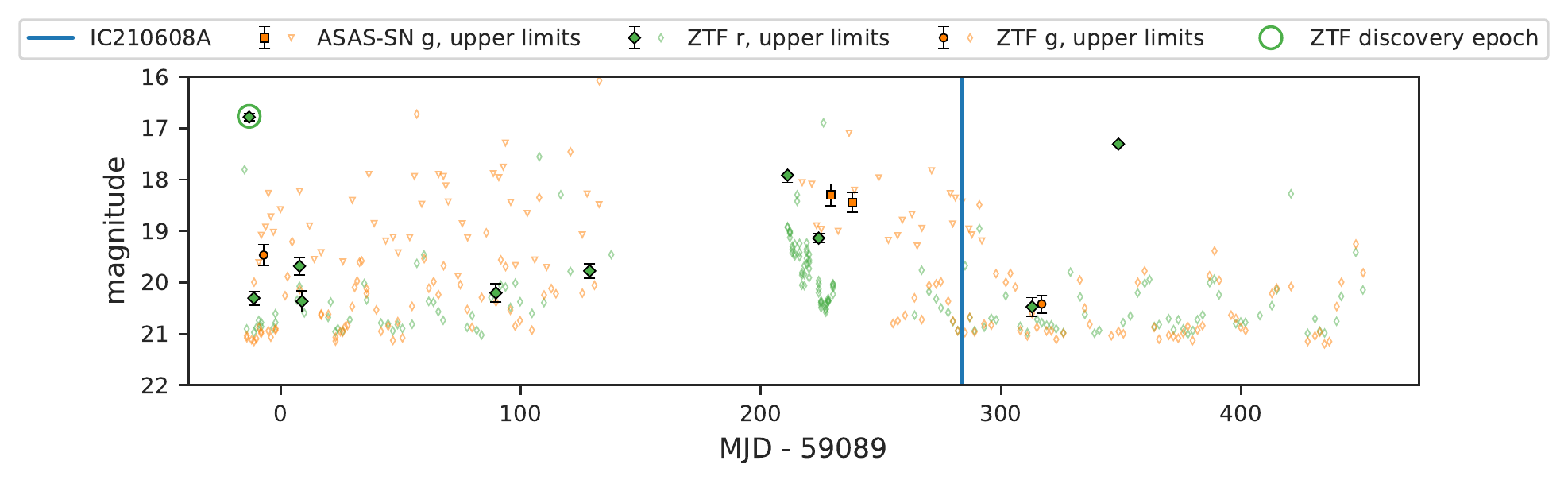}
\label{fig:at2020rng}
\end{center}

\section{}\label{AppendixB}

In Table \ref{tab:asassn_transients_long}, the relevant information for all the transients observed by ASAS-SN in IceCube regions is displayed. The first column gives the transient name, followed (in Column 2) by the epoch of the ASAS-SN discovery. In Column 3, we list the IceCube neutrino alert name while column 4 contains the ASAS-SN discovery time. Finally, in Column 5 we give the difference in days between the ASAS-SN and IceCube epochs and in columns 6 the transient type. Interesting sources as defined in Section \ref{sec:counterpart_classes} that were discovered within 500 days before the neutrino are discussed in more detail in Section \ref{sec:candidate_counterparts} while all other sources are not considered potential neutrino counterparts.
\footnotesize
\begin{longtable}{c r r r r c}
\caption{The transient sample.}\\
\hline
\hline 
\multicolumn{1}{c}{Transient} & \multicolumn{1}{c}{ASAS-SN detection} & \multicolumn{1}{c}{IceCube alert} & \multicolumn{1}{c}{Alert epoch} & \multicolumn{1}{c}{$\Delta_{\rm t}=t_{\rm ASASSN}-t_{\rm IceCube}$} & \multicolumn{1}{c}{Transient type} \\
\multicolumn{1}{c}{} & \multicolumn{1}{c}{JD} & \multicolumn{1}{c}{} & \multicolumn{1}{c}{JD} & \multicolumn{1}{c}{days} & \multicolumn{1}{c}{}\\
\hline
\endfirsthead

\hline
\hline 
\multicolumn{1}{c}{Transient} & \multicolumn{1}{c}{ASAS-SN discovery} & \multicolumn{1}{c}{IceCube alert} & \multicolumn{1}{c}{Alert epoch} & \multicolumn{1}{c}{$\Delta_{\rm t}=t_{\rm ASASSN}-t_{\rm IceCube}$} & \multicolumn{1}{c}{Transient type} \\
\multicolumn{1}{c}{} & \multicolumn{1}{c}{JD} & \multicolumn{1}{c}{} & \multicolumn{1}{c}{JD} & \multicolumn{1}{c}{days} & \multicolumn{1}{c}{}\\
\hline
\endhead

\hline 
\endlastfoot
SN~2021yrf &2459472.9 &IC200911A &2459104.1 &369 &SN~Ia\\
ASASSN-21qp &2459457.8 &IC200530A &2458999.8 &458 &SN~Ia\\
SN~2021vtl &2459442.0 &IC200614A &2459015.0 &427 &SN~Ia\\
SN~2021vpv &2459439.0 &IC210608A &2459373.7 &65 &SN~Ia\\
ASASSN-21oy &2459431.7 &IC200512A &2458981.8 &450 &CV\\
SN~2021ucx &2459430.1 &IC200614A &2459015.0 &415 &SN~Ia\\
SN~2021rem &2459405.7 &IC200410A &2458950.5 &455 &SN~II\\
SN~2021rgw &2459399.9 &IC200410A &2458950.5 &449 &SN~Ia\\
AT2021qiz &2459394.5 &IC160814A &2457615.4 &1779 &Unknown\\
ASASSN-21ld &2459380.1 &IC201120A &2459173.9 &206 &Flare\\
SN~2021oza &2459373.8 &IC200410A &2458950.5 &423 &SN~Ia\\
SN~2021mim &2459365.8 &IC200410A &2458950.5 &415 &SN~Ia\\
SN~2021mid &2459364.9 &IC200410A &2458950.5 &414 &SN~Ia\\
SN~2021jwl &2459329.9 &IC191119A &2458806.5 &523 &SN~Ia\\
ASASSN-21gk &2459322.9 &IC190221A &2458535.9 &787 &CV\\
SN~2021ipb &2459316.9 &IC200410A &2458950.5 &366 &SN~Ia\\
SN~2021hem &2459310.8 &IC200410A &2458950.5 &360 &SN~Ia\\
SN~2021ghc &2459307.8 &IC190819A &2458715.2 &593 &SN~Ia\\
SN~2021fqb &2459296.0 &IC191119A &2458806.5 &489 &SN~Ia\\
SN~2021ezt &2459293.0 &IC200530A &2458999.8 &293 &SN~IIn\\
AT2021fnf &2459290.7 &IC200410A &2458950.5 &340 &QSO\\
ASASSN-21dk &2459286.8 &IC200921A &2459114.3 &172 &CV\\
SN~2021bnw &2459262.9 &IC200109A &2458858.5 &404 &SLSN-I\\
AT2021cgr &2459257.5 &IC201130A &2459184.3 &73 &Unknown\\
ASASSN-21at &2459249.8 &IC190712A &2458676.6 &573 &CV\\
SN~2020aaxo &2459187.8 &IC200523A &2458992.6 &195 &SN~Ia\\
ASASSN-20pb &2459185.8 &IC200425A &2458965.5 &220 &CV\\
SN~2020yxd &2459170.8 &IC200614A &2459015.0 &156 &SN~Ia\\
SN~2020yfw &2459168.7 &IC210608A &2459373.7 &-205 &SN~Ia\\
SN~2020unl &2459138.1 &IC200425A &2458965.5 &173 &SN~Ia\\
ASASSN-20my &2459131.6 &IC190221A &2458535.9 &596 &CV\\
ASASSN-20mb &2459106.8 &IC200512A &2458981.8 &125 &CV\\
AT2020rng &2459089.9 &IC210608A &2459373.7 &-284 &Unknown\\
ASASSN-20jl &2459061.5 &IC200410A &2458950.5 &111 &CV\\
SN~2020oye &2459054.8 &IC200410A &2458950.5 &104 &SN~Ia\\
AT2020neh &2459024.8 &IC200410A &2458950.5 &74 &TDE\\
SN~2020lsc &2459011.9 &IC200410A &2458950.5 &61 &SN~Ia\\
SN~2020kpx &2458994.8 &IC191119A &2458806.5 &188 &SN~Ia\\
SN~2020joh &2458985.8 &IC200410A &2458950.5 &35 &SN~Ia\\
ASASSN-20eh &2458964.8 &IC200530A &2458999.8 &-35 &CV\\
AT2020idu &2458962.9 &IC200530A &2458999.8 &-37 &Unknown\\
ASASSN-20ea &2458957.7 &IC200916A &2459109.4 &-152 &CV\\
ASASSN-20dz &2458956.6 &IC201120A &2459173.9 &-217 &CV\\
ASASSN-20dy &2458955.9 &IC201221A &2459205.0 &-249 &CV\\
AT2020fhs &2458943.1 &IC200410A &2458950.5 &-7 &SN~Ia\\
SN~2020cli &2458909.9 &IC200410A &2458950.5 &-41 &SN~Ia\\
SN~2020czo &2458904.1 &IC200109A &2458858.5 &46 & SN~Ia\\
AT2020cxg &2458899.7 &IC160814A &2457615.4 &1284 &CV\\
SN~2020zj &2458876.0 &IC200410A &2458950.5 &-74 &SN~Ia\\
AT2020ajp &2458870.1 &IC160731A &2457600.6 &1270 &Unknown\\
ASASSN-20aq &2458867.2 &IC200921A &2459114.3 &-247 &SN~Ia\\
SN~2020ds &2458863.6 &IC201209A &2459192.9 &-329 &SN~Ia\\
SN~2019zhs &2458851.1 &IC191119A &2458806.5 &45 &SN~Ia\\
ASASSN-19aeb &2458845.7 &IC210608A &2459373.7 &-528 &SN~Ia\\
ASASSN-19abo &2458810.5 &IC190819A &2458715.2 &95 &SN~II\\
ASASSN-19abe &2458792.5 &IC200512A &2458981.8 &-189 &CV\\
SN~2019stx &2458786.5 &IC191122A &2458810.4 &-24 &SN~Ia\\
ASASSN-19aae &2458784.6 &IC210608A &2459373.7 &-589 &CV\\
ASASSN-19wi &2458730.8 &IC210608A &2459373.7 &-643 &CV\\
AT2019pfq &2458730.8 &IC210608A &2459373.7 &-643 &Unknown\\
SN~2019oml &2458721.6 &IC191231A &2458849.0 &-127 &SN~Ia\\
ASASSN-19uo &2458719.8 &IC200410A &2458950.5 &-231 &SN~Ia\\
ASASSN-19ua &2458715.5 &IC200523A &2458992.6 &-277 &CV.\\
AT2019lvs &2458697.4 &IC210608A &2459373.7 &-676 &Unknown\\
SN~2019lrc &2458689.0 &IC200523A &2458992.6 &-304 &SN~Ia\\
SN~2019kze &2458679.8 &IC210608A &2459373.7 &-694 &SN~Ia\\
SN~2019ieh &2458673.8 &IC200410A &2458950.5 &-277 &SN~Ic\\
SN~2019igg &2458666.9 &IC200410A &2458950.5 &-284 &SN~Ia\\
ASASSN-19qg &2458661.2 &IC200109A &2458858.5 &-197 &SN~Ia\\
ASASSN-19px &2458657.8 &IC200410A &2458950.5 &-293 &CV\\
SN~2019gqa &2458644.9 &IC200410A &2458950.5 &-306 &SN~Ia\\
AT2019fxr &2458634.9 &IC200410A &2458950.5 &-316 &Unknown\\
AT2019dsg &2458618.9 &IC191001A &2458758.3 &-139 &TDE\\
SN~2019due &2458601.8 &IC200410A &2458950.5 &-349 &SN~Ia\\
ASASSN-19kw &2458600.9 &IC190104A &2458487.9 &113 &SN~Ia\\
SN~2019crd &2458576.9 &IC190704A &2458669.3 &-92 &SN~Ia\\
SN~2019byw &2458572.6 &IC200410A &2458950.5 &-378 &SN~Ia\\
ASASSN-19hm &2458569.6 &IC160814A &2457615.4 &954 &SN~Ia\\
ASASSN-19dw &2458544.5 &IC190704A &2458669.3 &-125 &SN~Ia\\
ASASSN-19dz &2458545.0 &IC200410A &2458950.5 &-406 &SN~Ia\\
SN~2019aah &2458519.6 &IC191119A &2458806.5 &-287 &SN~II\\
ASASSN-19bx &2458516.0 &IC200410A &2458950.5 &-434 &SN~Ia\\
SN~2019agm &2458516.0 &IC200410A &2458950.5 &-434 &SN~Ia\\
SN~2019pe &2458502.0 &IC200410A &2458950.5 &-449 &SN~Ia\\
SN~2019vv &2458502.0 &IC191119A &2458806.5 &-305 &SN~Ia\\
ASASSN-19an &2458493.0 &IC200425A &2458965.5 &-472 &CV\\
SN~2018kji &2458484.0 &IC200410A &2458950.5 &-466 &SN~Ia\\
ASASSN-18abn &2458465.6 &IC191122A &2458810.4 &-345 &CV.\\
ASASSN-18abl &2458465.4 &IC191231A &2458849.0 &-384 &Stellar outburst\\
SN~2018ids &2458450.6 &IC191001A &2458758.3 &-308 &SN~Ia\\
SN~2018hom &2458423.8 &IC190619A &2458654.1 &-230 &SN~Ic-BL\\
SN~2018hhn &2458417.7 &IC190619A &2458654.1 &-236 &SN~Ia\\
SN~2018hcr &2458401.6 &IC200523A &2458992.6 &-591 &SN~Ia\\
ASASSN-18um &2458369.5 &IC210608A &2459373.7 &-1004 &CV\\
SN~2018fng &2458369.5 &IC200410A &2458950.5 &-581 &SN~Ia\\
AT2018fce &2458365.6 &IC200410A &2458950.5 &-585 &Unknown\\
ASASSN-18rm &2458339.7 &IC210608A &2459373.7 &-1034 &CV\\
SN~2018eoe &2458338.8 &IC210608A &2459373.7 &-1035 &SN~Ib\\
SN~2018emi &2458333.3 &IC200530A &2458999.8 &-667 &SN~Ia\\
ASASSN-18qo &2458331.9 &IC201120A &2459173.9 &-842 &CV\\
AT2018dzy &2458320.9 &IC210608A &2459373.7 &-1053 &SN~Ia\\
AT2018dyd &2458316.8 &IC191001A &2458758.3 &-442 &SN~Ia\\
SN~2018cyh &2458303.5 &IC191119A &2458806.5 &-503 &SN~II\\
ASASSN-18nr &2458289.0 &IC210608A &2459373.7 &-1085 &SN~Ia\\
ASASSN-18mx &2458286.1 &IC190619A &2458654.1 &-368 &SN~II\\
ASASSN-18mn &2458282.6 &IC190104A &2458487.9 &-205 &SN~Ia\\
AT2018chp &2458282.9 &IC200410A &2458950.5 &-668 &Unknown\\
SN~2018bwr &2458274.7 &IC200410A &2458950.5 &-676 &SN~IIn\\
ASASSN-18ll &2458271.5 &IC191119A &2458806.5 &-535 &Unknown\\
ASASSN-18kt &2458258.8 &IC210510A &2459344.7 &-1086 &CV\\
SN~2018bgy &2458250.8 &IC181014A &2458406.0 &-155 &SN~Ia\\
SN~2018bac &2458241.5 &IC190704A &2458669.3 &-428 &SN~Ia\\
ASASSN-18cr &2458162.9 &IC200921A &2459114.3 &-951 &CV\\
ASASSN-18bs &2458143.6 &IC190629A &2458664.3 &-521 &CV\\
SN~2018ie &2458140.8 &IC171015A &2458041.6 &99 &SN~Ic\\
SN~2017ivu &2458110.1 &IC200410A &2458950.5 &-840 &SN~II\\
ASASSN-17ot &2458070.8 &IC180908A &2458370.3 &-300 &Unknown\\
ASASSN-17ou &2458067.7 &IC190619A &2458654.1 &-586 &CV\\
ASASSN-17oc &2458059.7 &IC210608A &2459373.7 &-1314 &SN~Ia\\
SN~2017hfw &2458046.7 &IC201209A &2459192.9 &-1146 &SN~Ic\\
ASASSN-17od &2458046.1 &IC200615A &2459016.1 &-970 &CV\\
ASASSN-17lz &2458008.5 &IC200410A &2458950.5 &-942 &SN~Ia\\
AT2017fwf &2457977.8 &IC200410A &2458950.5 &-973 &Unknown\\
ASASSN-17jz &2457961.9 &IC201221A &2459205.0 &-1243 &ANT\\
ASASSN-17jt &2457960.1 &IC191001A &2458758.3 &-798 &Unknown\\
ASASSN-17ji &2457946.5 &IC191001A &2458758.3 &-812 &CV\\
ASASSN-17ja &2457935.5 &IC200410A &2458950.5 &-1015 &CV\\
ASASSN-17ii &2457927.9 &IC200410A &2458950.5 &-1023 &SN~Ia\\
ASASSN-17gr &2457897.9 &IC200410A &2458950.5 &-1053 &Unknown\\
AT2017edw &2457893.8 &IC200109A &2458858.5 &-965 & Unknown\\
AT2017dtj &2457891.9 &IC191119A &2458806.5 &-915 &Unknown\\
ASASSN-17ff &2457863.8 &IC200109A &2458858.5 &-995 &SN~Ia\\
ASASSN-17fh &2457864.0 &IC200410A &2458950.5 &-1086 &CV.\\
ASASSN-17fd &2457863.0 &IC191119A &2458806.5 &-944 &SN~Ia\\
ASASSN-17fd &2457863.0 &IC200410A &2458950.5 &-1087 &SN~Ia\\
SN~2017cne &2457847.0 &IC191119A &2458806.5 &-960 &SN~Ia\\
ASASSN-17ei &2457844.8 &IC181014A &2458406.0 &-561 &CV\\
SN~2017ciy &2457840.0 &IC200530A &2458999.8 &-1160 &SN~Ia\\
ASASSN-17eb &2457835.6 &IC200410A &2458950.5 &-1115 &SN~Ia\\
ASASSN-17cr &2457805.1 &IC200410A &2458950.5 &-1145 &SN~Ia\\
SN~2017avl &2457802.8 &IC191204A &2458822.4 &-1020 &SN~Ia\\
AT2017jn &2457782.1 &IC200410A &2458950.5 &-1168 &Unknown\\
ASASSN-17bd &2457777.1 &IC200410A &2458950.5 &-1173 &SN~Ia\\
ASASSN-17bb &2457777.1 &IC191119A &2458806.5 &-1029 &SN~Ia\\
ASASSN-17ae &2457758.1 &IC200410A &2458950.5 &-1192 &SN~Ia\\
ASASSN-16ps &2457753.7 &IC190503A &2458607.2 &-854 &CV\\
ASASSN-16oo &2457728.6 &IC191122A &2458810.4 &-1082 &SN~Ia\\
AT2016ipd &2457721.9 &IC200614A &2459015.0 &-1293 &Unknown\\
SN~2016hvu &2457702.8 &IC210608A &2459373.7 &-1671 &SN~II\\
ASASSN-16ie &2457607.8 &IC200410A &2458950.5 &-1343 &SN~Ia\\
ASASSN-16hz &2457601.8 &IC200523A &2458992.6 &-1391 &SN~Ia\\
ASASSN-16fp &2457536.0 &IC210608A &2459373.7 &-1838 &SN~Ic-BL\\
ASASSN-16ex &2457511.9 &IC200530A &2458999.8 &-1488 &SN~Ia\\
SN~2016brw &2457511.9 &IC200530A &2458999.8 &-1488 &SN~II\\
MASTER OT J152333.04+092125.6 &2457489.9 &IC200410A &2458950.5 &-1461 &SN~Ia\\
ASASSN-16eg &2457488.0 &IC201221A &2459205.0 &-1717 &CV\\
ASASSN-16dp &2457479.0 &IC191119A &2458806.5 &-1328 &SN~Ia\\
ASASSN-16ct &2457454.0 &IC191119A &2458806.5 &-1353 &SN~Ia\\
ASASSN-16cq &2457453.8 &IC190712A &2458676.6 &-1223 &CV\\
SN~2016afa &2457450.0 &IC200410A &2458950.5 &-1500 &SN~II\\
ASASSN-16bg &2457425.0 &IC200921A &2459114.3 &-1689 &SN~Ia\\
MASTER OT J105908.57+103834.8 &2457419.0 &IC200109A &2458858.5 &-1440 &SN~Ia\\
PSNJ01534240+2956107 &2457386.8 &IC200614A &2459015.0 &-1628 &SN\\
ASASSN-15ul &2457380.1 &IC191119A &2458806.5 &-1426 &SN~Ia\\
ASASSN-15tp &2457360.5 &IC190712A &2458676.6 &-1316 &CV\\
ASASSN-15ti &2457357.9 &IC200911A &2459104.1 &-1746 &SN~Ia\\
PSNJ23002463+0137368 &2457247.0 &IC200523A &2458992.6 &-1746 &SN~Ib\\
ASASSN-15nz &2457242.8 &IC200410A &2458950.5 &-1708 &CV\\
ASASSN-15nr &2457240.9 &IC201021A &2459143.8 &-1903 &SN~Ia\\
ASASSN-15mu &2457222.0 &IC190619A &2458654.1 &-1432 &CV\\
ASASSN-15mw &2457220.5 &IC160814A &2457615.4 &-395 &blazar candidate\\
ASASSN-15js &2457163.8 &IC200109A &2458858.5 &-1695 &SN~Ia\\
ASASSN-15jd &2457156.0 &IC200410A &2458950.5 &-1795 &CV\\
ASASSN-15fu &2457108.7 &IC190922A &2458748.9 &-1640 &CV\\
PSNJ15053007+0138024 &2457101.1 &IC191119A &2458806.5 &-1705 &SN~Ia\\
ASASSN-15fm &2457096.6 &IC210510A &2459344.7 &-2248 &CV\\
ASASSN-15dz &2457074.1 &IC200410A &2458950.5 &-1876 &SN~Ia\\
ASASSN-15db &2457069.1 &IC200410A &2458950.5 &-1881 &SN~Ia\\
ASASSN-15dd &2457069.1 &IC200410A &2458950.5 &-1881 &SN~Ia\\
ASASSN-15bk &2457042.1 &IC200410A &2458950.5 &-1908 &SN~Ia\\
ASASSN-15bd &2457040.1 &IC200410A &2458950.5 &-1910 &SN~IIb\\
ASASSN-15av &2457036.9 &IC191204A &2458822.4 &-1786 &CV\\
PSNJ03281419+3801111 &2457018.8 &IC200911A &2459104.1 &-2085 &SN\\
2014ea &2457013.0 &IC190704A &2458669.3 &-1656 &SN~Ia\\
ASASSN-14im &2456932.0 &IC200911A &2459104.1 &-2172 &CV\\
ASASSN-14hi &2456916.5 &IC191215A &2458833.0 &-1916 &CV\\
PYPer &2456910.0 &IC200911A &2459104.1 &-2194 &CV\\
V589Her &2456904.8 &IC200410A &2458950.5 &-2046 &CV\\
ASASSN-14fq &2456885.6 &IC191204A &2458822.4 &-1937 &CV\\
CSS111118:051923+155435 &2456884.6 &IC190712A &2458676.6 &-1792 &CV\\
HYPsc &2456867.0 &IC200523A &2458992.6 &-2126 &CV\\
V544Her &2456857.9 &IC200410A &2458950.5 &-2093 &CV\\
ASASSN-14dc &2456833.1 &IC200614A &2459015.0 &-2182 &SN\\
SDSSJ102637.04+475426.3 &2456824.7 &IC200806A &2459068.1 &-2243 &CV\\
ASASSN-14ax &2456782.0 &IC200530A &2458999.8 &-2218 &SN~Ia\\
ASASSN-14an &2456758.0 &IC210811A &2459437.6 &-2680 &Nova\\
SDSSJ162520.29+120308.7 &2456757.9 &IC200410A &2458950.5 &-2193 &CV\\
2QZJ130441.7+010330 &2456736.8 &IC201115A &2459168.6 &-2432 &CV\\
ASASSN-13dg &2456567.9 &IC200614A &2459015.0 &-2447 &CV.\\
ASASSN-13cw &2456543.9 &IC191001A &2458758.3 &-2214 &CV\\
HS0218+3229 &2456540.9 &IC200614A &2459015.0 &-2474 &CV\\
1RXSJ185310.0+594509 &2456519.8 &IC191215A &2458833.0 &-2313 &CV\\
ASASJ224349+0809.5 &2456518.9 &IC190619A &2458654.1 &-2135 &CV\\
ASASSN-13bx &2456511.9 &IC200523A &2458992.6 &-2481 &CV\\
ASASSN-13bw &2456509.5 &IC200410A &2458950.5 &-2441 &CV\\
V521Peg &2456507.0 &IC210608A &2459373.7 &-2867 &CV\\
CSS110921:160824+165240 &2456482.8 &IC200410A &2458950.5 &-2468 &CV\\
ASASSN-13bi &2456482.8 &IC191119A &2458806.5 &-2324 &Flare\\
WYTri &2456481.1 &IC200614A &2459015.0 &-2534 &CV\\
QWSer &2456477.8 &IC200410A &2458950.5 &-2473 &CV\\
V368Peg &2456468.5 &IC190619A &2458654.1 &-2186 &CV\\
CSS080505:163121+103134 &2456468.9 &IC200410A &2458950.5 &-2482 &CV\\
CSS090911:221232+160140 &2456459.9 &IC210608A &2459373.7 &-2914 &CV\\

\hline
\label{tab:asassn_transients_long}
\end{longtable}
\setlength\LTleft{+0.8cm}
\vspace{1cm}

\bsp	
\label{lastpage}
\end{document}

%% file: latex/alert_table_observed.tex
IC160427A & 240.57 & $+9.34$ & 1.4 & 100.0 & 100.0 & - & \cite{ic160427a} \\ 
IC160731A & 214.50 & $-0.33$ & 2.2 & 36.2 & 100.0 & 85 & - \\ 
IC160814A & 200.30 & $-32.40$ & 12.0 & 0.0 & 100.0 & - & \cite{ic160814a} \\ 
IC161103A & 40.83 & $+12.56$ & 3.1 & 79.9 & 100.0 & - & \cite{ic161103a} \\ 
IC161210A & 46.58 & $+14.98$ & 1.7 & 0.0 & 100.0 & 49 & \cite{ic161210a} \\ 
IC170312A & 305.15 & $-26.61$ & 0.9 & 0.0 & 100.0 & - & \cite{ic170312a} \\ 
IC170321A & 98.30 & $-15.02$ & 5.6 & 4.5 & 100.0 & 28 & \cite{ic170321a} \\ 
IC170922A & 77.43 & $+5.72$ & 1.3 & 100.0 & 100.0 & 57 & \cite{ic170922a} \\ 
IC171106A & 340.00 & $+7.40$ & 0.7 & 100.0 & 100.0 & 75 & \cite{ic171106a} \\ 
IC181023A & 270.18 & $-8.57$ & 9.3 & 70.5 & 100.0 & 28 & \cite{ic181023a} \\ 
IC190104A & 357.98 & $-26.65$ & 18.5 & 14.0 & 100.0 & - & \cite{ic190104a} \\ 
IC190221A & 268.81 & $-17.04$ & 5.2 & 78.6 & 100.0 & - & \cite{ic190221a} \\ 
IC190331A & 337.68 & $-20.70$ & 0.4 & 0.0 & 100.0 & - & \cite{ic190331a} \\ 
IC190503A & 120.28 & $+6.35$ & 1.9 & 100.0 & 100.0 & 36 & \cite{ic190503a} \\ 
IC190619A & 343.26 & $+10.73$ & 27.2 & 100.0 & 100.0 & 55 & \cite{ic190619a} \\ 
IC190629A & 27.22 & $+84.33$ & 5.0 & 0.0 & 70.6 & 34 & \cite{ic190629a} \\ 
IC190704A & 161.85 & $+27.11$ & 21.0 & 100.0 & 100.0 & 49 & \cite{ic190704a} \\ 
IC190712A & 76.46 & $+13.06$ & 92.0 & 0.0 & 13.1 & 30 & \cite{ic190712a} \\ 
IC190730A & 225.79 & $+10.47$ & 5.4 & 100.0 & 100.0 & 67 & \cite{ic190730a} \\ 
IC190922B & 5.76 & $-1.57$ & 4.5 & 100.0 & 100.0 & 51 & \cite{ic190922b} \\ 
IC191001A & 314.08 & $+12.94$ & 25.5 & 100.0 & 100.0 & 59 & \cite{ic191001a} \\ 
IC191122A & 27.25 & $-0.04$ & 12.2 & 100.0 & 100.0 & 33 & \cite{ic191122a} \\ 
IC191204A & 79.72 & $+2.80$ & 11.6 & 98.8 & 100.0 & 33 & \cite{ic191204a} \\ 
IC191215A & 285.87 & $+58.92$ & 12.8 & 0.0 & 12.4 & 47 & \cite{ic191215a} \\ 
IC191231A & 46.36 & $+20.42$ & 35.6 & 100.0 & 100.0 & 46 & \cite{ic191231a} \\ 
IC200107A & 148.18 & $+35.46$ & 7.6 & 0.0 & 78.2 & 50$^{*}$ & \cite{ic200107a} \\ 
IC200109A & 164.49 & $+11.87$ & 22.5 & 77.7 & 100.0 & 77 & \cite{ic200109a} \\ 
IC200117A & 116.24 & $+29.14$ & 2.9 & 0.0 & 100.0 & 38 & \cite{ic200117a} \\ 
IC200410A & 242.58 & $+11.61$ & 377.9 & 38.0 & 100.0 & 31 & \cite{ic200410a} \\ 
IC200425A & 100.10 & $+53.57$ & 18.8 & 7.2 & 100.0 & 48 & \cite{ic200425a} \\ 
IC200512A & 295.18 & $+15.79$ & 9.8 & 62.5 & 100.0 & 32 & \cite{ic200512a} \\ 
IC200523A & 338.64 & $+1.75$ & 90.6 & 24.5 & 100.0 & 25 & \cite{ic200523a} \\ 
IC200530A & 255.37 & $+26.61$ & 25.3 & 92.4 & 100.0 & 59 & \cite{ic200530a} \\ 
IC200614A & 33.84 & $+31.61$ & 47.8 & 35.2 & 100.0 & 42 & \cite{ic200614a} \\ 
IC200615A & 142.95 & $+3.66$ & 5.9 & 97.9 & 100.0 & 83 & \cite{ic200615a} \\ 
IC200620A & 162.11 & $+11.95$ & 1.7 & 100.0 & 100.0 & 32 & \cite{ic200620a} \\ 
IC200911A & 51.11 & $+38.11$ & 52.7 & 46.5 & 100.0 & 41 & \cite{ic200911a} \\ 
IC200916A & 109.78 & $+14.36$ & 4.2 & 100.0 & 100.0 & 32 & \cite{ic200916a} \\ 
IC200926A & 96.46 & $-4.33$ & 1.7 & 100.0 & 100.0 & 44 & \cite{ic200926a} \\ 
IC200929A & 29.53 & $+3.47$ & 1.1 & 65.1 & 100.0 & 47 & \cite{ic200929a} \\ 
IC201007A & 265.17 & $+5.34$ & 0.6 & 0.0 & 100.0 & 88 & \cite{ic201007a} \\ 
IC201021A & 260.82 & $+14.55$ & 6.9 & 2.6 & 100.0 & 30 & \cite{ic201021a} \\ 
IC201114A & 105.25 & $+6.05$ & 4.5 & 100.0 & 100.0 & 56 & \cite{ic201114a} \\ 
IC201115A & 195.12 & $+1.38$ & 6.6 & 0.0 & 100.0 & 46 & \cite{ic201115a} \\ 
IC201120A & 307.53 & $+40.77$ & 64.3 & 82.5 & 100.0 & 50 & \cite{ic201120a} \\ 
IC201130A & 30.54 & $-12.10$ & 5.4 & 100.0 & 100.0 & 15 & \cite{ic201130a} \\ 
IC201209A & 6.86 & $-9.25$ & 4.7 & 100.0 & 100.0 & 19 & \cite{ic201209a} \\ 
IC201221A & 261.69 & $+41.81$ & 8.9 & 0.0 & 100.0 & 56 & \cite{ic201221a} \\ 
IC201222A & 206.37 & $+13.44$ & 1.5 & 100.0 & 100.0 & 53 & \cite{ic201222a} \\ 
IC210210A & 206.06 & $+4.78$ & 2.8 & 100.0 & 100.0 & 65 & \cite{ic210210a} \\ 
IC210503A & 143.53 & $+41.81$ & 102.6 & 27.1 & 100.0 & 41 & \cite{ic210503a} \\ 
IC210510A & 268.42 & $+3.81$ & 4.0 & 0.0 & 100.0 & 28 & \cite{ic210510a} \\ 
IC210608A & 337.41 & $+18.37$ & 109.7 & 94.8 & 100.0 & 31 & \cite{ic210608a} \\ 
IC210629A & 340.75 & $+12.94$ & 6.0 & 100.0 & 100.0 & 35 & \cite{ic210629a} \\ 
IC210717A & 46.49 & $-1.34$ & 30.0 & 69.2 & 100.0 & 50$^{*}$ & \cite{ic210717a} \\ 
IC210811A & 270.79 & $+25.28$ & 3.2 & 100.0 & 100.0 & 66 & \cite{ic210811a} \\ 

%% file: latex/alert_table_not_observed.tex
	 IC160806A & 122.81 & $-0.81$ & 0.0 & proximity to sun & \cite{ic160806a} \\ 
	 IC171015A & 162.86 & $-15.44$ & 14.9 & proximity to sun & \cite{ic171015a} \\ 
	 IC180908A & 144.58 & $-2.13$ & 6.3 & proximity to sun & \cite{ic180908a} \\ 
	 IC181014A & 225.15 & $-34.80$ & 10.5 & proximity to sun & \cite{ic181014a} \\ 
	 IC190124A & 307.4 & $-32.18$ & 2.0 & proximity to sun & \cite{ic190124a} \\ 
	 IC190819A & 148.8 & $+1.38$ & 9.3 & proximity to sun & \cite{ic190819a} \\ 
	 IC190922A & 167.43 & $-22.39$ & 32.2 & proximity to sun & \cite{ic190922a} \\ 
	 IC191119A & 230.1 & $+3.17$ & 61.2 & proximity to sun & \cite{ic191119a} \\ 
	 IC200421A & 87.93 & $+8.23$ & 24.4 & operation & \cite{ic200421a} \\ 
	 IC200806A & 157.25 & $+47.75$ & 1.8 & proximity to sun & \cite{ic200806a} \\ 
	 IC200921A & 195.29 & $+26.24$ & 12.0 & proximity to sun & \cite{ic200921a} \\ 
	 IC200926B & 184.75 & $+32.93$ & 9.0 & proximity to sun & \cite{ic200926b} \\ 
	 IC201014A & 221.22 & $+14.44$ & 1.9 & proximity to sun & \cite{ic201014a} \\ 
	 IC210516A & 91.76 & $+9.52$ & 2.2 & proximity to sun & \cite{ic210516a} \\ 
	 IC210730A & 105.73 & $+14.79$ & 6.6 & proximity to sun & \cite{ic210730a} \\ 